\documentclass[aps, preprint, review, amsmath,amssymb]{revtex4-2}

 \usepackage{graphics, subfigure}
 \usepackage{graphicx}	

\usepackage{amssymb, amsmath, mathtools}

\usepackage{hyperref}
\hypersetup{colorlinks = true, 
  urlcolor   = blue, 
  linkcolor  = cyan, 
  citecolor  = blue} 

\hypersetup{pdfstartview=FitH}
\hypersetup{pdfpagemode=UseNone}

\usepackage{bm}

\usepackage{comment}

\usepackage{color}

\usepackage{upgreek} 

\renewcommand{\Re}{\mbox{Re}}
  \newcommand{\ea}{\bm{e}_\alpha}
  \newcommand{\eb}{\bm{e}_\beta}

\newcommand{\ga}{g_{\alpha}}
\newcommand{\gaeq}{g_{\alpha}^{\mbox{\tiny eq}}}

\newcommand{\Fb}{\mathbf{F}_{\rm b}}
\newcommand{\uw}{\bm{u}_{\rm w}}

\newcommand{\etal}{\textit{et al.}}

\begin{document}

\preprint{arXiv}

\title{\textbf{Improving the accuracy of the momentum-exchange method in lattice Boltzmann methods}}

\author{Abbas Fakhari}
\email{afakhari@odu.edu}
\affiliation{Department of Mathematics and Statistics, Old Dominion University, Norfolk, VA 23508, USA}

\date{\today}

\begin{abstract}

In the lattice Boltzmann method (LBM), the momentum-exchange method (MEM) is widely used to evaluate hydrodynamic forces at fluid-solid interfaces due to its computational efficiency and geometric flexibility. However, when mapping complex geometries onto structured Cartesian grids, the physical boundary rarely intersects the lattice links exactly at their midpoints. This discrepancy violates an implicit assumption inherent to the conventional momentum-exchange (CME) formulation. In this study, we systematically analyze this geometric deficiency and demonstrate that it induces significant force inaccuracies, even in elementary flows where the solid boundary deviates from the midway position. To resolve this, we propose a modified momentum-exchange (MME) method that explicitly accounts for the exact boundary location along the lattice links, demonstrating enhanced accuracy for both flat and curved geometries. 

Furthermore, we address another critical issue with the interpolation-supplemented bounce-back (ISBB) schemes, and that is their inherent violation of local mass conservation. This mass leakage induces unphysical pressure buildup in the system, which compromises the accuracy and stability of the LBM. We propose a mass-conservative remedy for ISBB schemes that eliminates this leakage. Through a series of benchmark simulations of increasing complexity, we demonstrate that combining the proposed MME with the mass-corrected ISBB scheme drastically improves the fidelity of hydrodynamic force evaluations, establishing a robust framework essential for moving boundary problems and fluid-structure interactions.

\end{abstract}


\maketitle

\newpage

\section{Introduction}
\label{introduction}

The Lattice Boltzmann Method (LBM) has emerged as a powerful mesoscopic numerical framework for simulating complex fluid flows and fluid-structure interactions (FSI) \cite{ChenDoolen1998, Aidun2010}. Unlike conventional Navier–Stokes solvers that rely on the discretization of macroscopic partial differential equations, the LBM evolves a discrete set of particle distribution functions (PDFs) on a regular Cartesian lattice. The core algorithm alternates between local collision and linear streaming steps, a decoupling that renders the method highly parallelizable and naturally second-order accurate in both space and time on uniform grids \cite{HL1997PRE, HL1997JSP}. This underlying Cartesian framework makes the LBM exceptionally efficient at handling complex, arbitrarily shaped solid geometries without the need for dynamic body-fitted mesh generation.

However, resolving non-body-aligned physical boundaries on a rigid Cartesian lattice introduces distinct algorithmic challenges. The standard approach for straight, axis-aligned walls is the halfway bounce-back (HWBB) scheme \cite{Ladd1994}. In the HWBB method, outgoing PDFs striking a solid boundary are reflected back along their incoming links toward the fluid nodes from which they originated. This reflection is implicitly assumed to occur exactly at the midpoint of the lattice link. The HWBB scheme is widely favored because it strictly conserves mass locally, preserves second-order spatial accuracy for grid-aligned walls, and is trivial to implement. 

To evaluate the hydrodynamic forces exerted by the fluid on embedded solid structures, Ladd \cite{Ladd1994} introduced the momentum-exchange method (MEM). Compared to the conventional stress-integration method, which requires computing velocity gradients and taking derivatives of the pressure field along complex surfaces, the MEM evaluates forces by summing the net momentum transferred by the PDFs as they bounce off the solid boundary. This localized algebraic summation preserves the computational efficiency of the LBM and scales seamlessly to complex particulate suspensions. 

Despite its advantages, the conventional momentum-exchange (CME) method possesses known theoretical limitations. Early formulations lacked Galilean invariance for moving boundaries, prompting several modifications to account for the momentum of the moving solid wall \cite{Aidun2010, Caiazzo2008, HuLiShuNiu2015, Peng2017PRE}. Notably, Wen \etal \cite{Wen2014} resolved these inconsistencies by introducing a Galilean-invariant momentum-exchange formulation. 


However, a more fundamental geometric issue persists when MEM is applied to non-axis-aligned or curved boundaries. To maintain second-order accuracy without resorting to a crude staircase grid approximation, the LBM is frequently coupled with interpolation-supplemented bounce-back (ISBB) schemes \cite{Bouzidi2001, Yu2003AIAA, LallemandLuo2003JCP, Zhao2017}. These schemes utilize linear or quadratic interpolations of the PDFs based on the exact fraction of the link, $q$, cut by the physical boundary. Crucially, the CME assumes the force exchange occurs precisely at the midpoint ($q=\frac{1}{2}$). When $q \neq \frac{1}{2}$, evaluating the force exchange without accounting for the true spatial location of the boundary introduces non-negligible errors. In this paper, we challenge this midlink assumption. We demonstrate analytically and numerically that when the physical boundary deviates from the midpoint, the standard MEM formulation must be adjusted to account for the localized link geometry to yield accurate force predictions.

A secondary, yet equally severe, challenge arises from the boundary condition itself. Although ISBB schemes successfully restore spatial accuracy for curved walls, they inherently violate strict mass conservation due to the asymmetric nature of the interpolation steps \cite{Bao2008JCP, Sanjeevi2018PRE}. This localized mass leakage leads to unphysical pressure fluctuations or cumulative pressure buildup within the computational domain. Because the MEM is intrinsically sensitive to the local pressure field, this mass-conservation error heavily degrades the accuracy and numerical stability of force evaluations at the fluid-solid interface. This issue is particularly exacerbated in closed systems where mass leakage errors accumulate.

To address these coupled deficiencies, this work presents a two-fold advancement for fluid-solid interfaces in the LBM. First, we propose a modified momentum-exchange (MME) method that corrects for arbitrary boundary locations along lattice links, validating its performance on both flat, inclined walls and curved boundaries. Second, we introduce a mass-conservative remedy for ISBB schemes that eliminates unphysical mass leakage and stabilizes the local pressure field. Through a systematic suite of benchmark problems of increasing complexity —beginning with a simple gravity-driven Poiseuille flow— we show that integrating the proposed MME with a mass-corrected ISBB scheme substantially improves the fidelity of hydrodynamic force calculations. This combined approach offers a highly accurate and stable framework tailored for complex fluid-structure interaction simulations.

The remainder of this paper is organized as follows: Section~\ref{incompressible_LBE} details the implementation of the nearly incompressible version of the LBM \cite{HL1997JSP}, followed by the mathematical formulation of the proposed MME and the mass-conservative ISBB scheme in Section~\ref{implementation}. Section~\ref{results} presents numerical validations and comparative studies against analytical solutions and existing literature, while Section~\ref{conclusion} outlines our concluding remarks and future perspectives.

\section{Nearly incompressible LBE}
\label{incompressible_LBE}

In this study, we implement the nearly incompressible formulation of the lattice Boltzmann equation (LBE) \cite{HL1997JSP}. The evolution equation for the pressure-based particle distribution function, $\ga$, is expressed as:
\begin{equation}
    \ga(\bm{x}+\ea \delta_t,t+\delta_t) = \ga(\bm{x},t) + \Omega_\alpha(\bm{x},t) + F_\alpha(\bm{x},t)
\label{LBE_g}
\end{equation}
where $\bm{x}$ represents the spatial coordinate, $t$ is time, $\delta_t$ is the discrete time step, $\ea$ is the set of mesoscopic velocities, $\Omega_\alpha$ is the collision operator, and $F_\alpha$ is the forcing term due to an external body force.
To enhance numerical stability and independently tune transport coefficients, a multiple-relaxation-time (MRT) collision model is implemented \cite{LallemandLuo2000} (detailed in Sec.~\ref{sec_collision}):
\begin{equation}
    \Omega_\alpha = - \mathbf{M}^{-1} \hat{\mathbf{S}} \mathbf{M} \left( \ga - \gaeq \right)
\label{eq_Omega}
\end{equation}
where $\mathbf{M}$ is the orthogonal transformation matrix mapping the PDFs into moment space, and $\hat{\mathbf{S}}$ is the diagonal relaxation matrix. The modified equilibrium distribution function (EDF) is defined as
\begin{equation}
    \gaeq = \rho c_s^2 \Gamma_\alpha + (p-\rho c_s^2) w_\alpha - \frac{1}{2} F_\alpha
\label{eq_geq}
\end{equation}
where $p$ is the macroscopic pressure, $\rho$ is the fluid density, $w_\alpha$ denotes the weighting factors associated with the discrete velocity directions, and $c_s = c/ \sqrt{3}$ is the lattice speed of sound. Here, $c = \delta_x / \delta_t$ represents the lattice speed, where $\delta_x$ is the lattice grid spacing. For standard uniform Cartesian lattices $\delta_x = \delta_t = c = 1$. The dimensionless velocity-dependent function $\Gamma_\alpha$ is given by:
\begin{equation}
    \Gamma_\alpha = w_\alpha 
          \left[
            1 + \frac{\ea \cdot \bm{u}}{c_s^2}
            + \frac{\left( \ea \cdot \bm{u} \right)^2}{2 c_s^4} - \frac{\bm{u \cdot u}}{2 c_s^2}
          \right]
\label{eq_feq}
\end{equation}
where $\bm{u}$ is the macroscopic velocity vector. For the D2Q9 lattice framework adopted throughout this work, the directional weights are given by $w_0=4/9$, $w_{1-4}=1/9$, and $w_{5-8}=1/36$ \cite{HL1997PRE}, with the discrete velocity vectors defined as:
\begin{equation}
    \ea = c
        \left\{
	       \begin{array}{lll}
    	       (0,0), & & \alpha = 0 \\
    	       (\cos\theta_\alpha,\sin\theta_\alpha), & \theta_\alpha = (\alpha-1)\pi/2, &\alpha=1 \mbox{\,—\,} 4 \\
    	       (\cos\theta_\alpha,\sin\theta_\alpha)\sqrt{2}, &\theta_\alpha=(2\alpha-9)\pi/4, & \alpha=5 \mbox{\,—\,} 8
	       \end{array}
	\right.
\end{equation}

To account for external body forces, a generalized discrete forcing term is incorporated using the framework introduced by He et al. \cite{HCZ1999}:
\begin{equation}
    F_\alpha(\bm{x},t) = \delta_t \Gamma_\alpha (\ea - \bm{u}) \cdot \Fb
\label{eq_Fa}
\end{equation}
where $\Fb = \rho\mathbf{G}$ represents the volumetric body force vector, and $\mathbf{G}$ is the macroscopic acceleration field. In the low-Mach-number hydrodynamic limit, the forcing term in Eq.~\eqref{eq_Fa} can be simplified by truncating higher-order velocity terms, reducing to:
\begin{equation}
    F_\alpha(\bm{x},t) = \delta_t w_\alpha \left[ ({\ea - \bm{u}}) + \frac{(\ea \cdot \bm{u})}{c_s^2} \ea \right] \cdot \Fb
\end{equation}
This is the same forcing term that was proposed by Guo et al. \cite{Guo2002}.

The nearly incompressible LBE formulation detailed in Eq.~\eqref{LBE_g} demonstrates superior performance in suppressing unphysical acoustic pressure waves and improving temporal accuracy during transient simulations compared to standard density-based formulations. This enhancement stems from the minimization of the inherent compressibility error, particularly in low-Mach-number unsteady regimes \cite{HL1997JSP}. When integrated with an MRT collision operator, this formulation establishes a robust and high-fidelity platform for simulating transient, high-Reynolds-number incompressible flows.

Following the execution of the decoupled collision and streaming phases, the macroscopic pressure and velocity fields are updated locally via the moments of the PDFs:
\begin{subequations}
\begin{align}
    p &= \sum_\alpha \ga
    \\
    \rho c_s^2 \bm{u} &= \sum_\alpha \ga \ea + \frac{\delta_t}{2} c_s^2 \Fb
\end{align}
\end{subequations}

\subsection{Macroscopic Hydrodynamic Equations}
\label{sec_NSE}

Through a formal Chapman–Enskog expansion, the nearly incompressible LBE recovers the following Navier–Stokes equations for isothermal, incompressible fluid flows up to second-order accuracy in space and time:
\begin{subequations}
\begin{align}
    \partial_t p + \rho c_s^2 \bm{\nabla} \cdot \bm{u}
    & = 0
    \\
    \rho \left( \partial_t \bm{u} +\bm{u} \cdot \bm{\nabla} \bm{u} \right)
    & = - \bm{\nabla} p + \bm{\nabla} \cdot \left( \mu \left[ \bm{\nabla}\bm{u} + (\bm{\nabla}\bm{u})^T \right] \right) + \Fb
\end{align}
\end{subequations}
in which $\mu$ is the dynamic viscosity.

\subsection{Collision Operator}
\label{sec_collision}

The multiple-relaxation-time (MRT) collision framework provides significantly enhanced numerical stability and accuracy compared to the standard single-relaxation-time (SRT) or Bhatnagar-Gross-Krook (BGK) approximations, particularly in high-Reynolds-number regimes \cite{LallemandLuo2000}. By decoupling the relaxation processes of distinct physical modes, the MRT model overcomes the severe numerical instabilities that typically arise in BGK models as the dimensionless relaxation time $\tau$ approaches $0$.

In the MRT framework defined in Eq.~\eqref{eq_Omega}, the orthogonal transformation matrix $\mathbf{M}$ maps the distribution functions from physical space to moment space via $\bm{m} = \mathbf{M}\bm{f}$, and $\mathbf{M}^{-1}$ defines its inverse mapping. The diagonal relaxation matrix $\hat{\mathbf{S}}$ dictating the rates at which individual moments relax toward their equilibria is configured as \cite{PENG2016CAMWA}:
\begin{equation}
    \hat{\mathbf{S}} = \text{diag}(1,\, s_\nu,\, 1,\, 1,\, s_b,\, 1,\, s_b,\, s_\nu,\, s_\nu)
\label{eq_S}
\end{equation}
where the non-trivial relaxation rates are determined by \cite{LallemandLuo2000}
\begin{subequations}
\begin{align}
    s_\nu &= \frac{1}{\tau+\frac{1}{2}} \\
    s_b   &= 8\frac{2-s_\nu}{8-s_\nu}
\end{align}
\end{subequations}
in which $\tau$ is related to the kinematic viscosity $\nu$ by
\begin{equation}
    \nu = \tau c_s^2 \delta_t
\end{equation}
Setting all diagonal relaxation parameters in $\hat{\mathbf{S}}$ identical to $s_\nu$ formally reduces the MRT operator to the standard BGK model.

In addition to handling numerical instability, the choice of collision model impacts boundary accuracy. For mid-link no-slip boundaries, standard BGK model exhibits a known coupling between the physical viscosity and the numerical location of the hydrodynamic wall, often requiring a specific "magic" relaxation parameter to fix the wall location \cite{GinzburgdHumieres2003}. This dependency severely restricts grid flexibility and limits the user's ability to freely adjust viscosity. Utilizing the MRT framework decouples these modes, permitting independent optimization of the bulk and shear relaxation rates, thereby ensuring that the no-slip boundary condition remains accurately positioned at the intended physical interface without compromising stability or accuracy \cite{LallemandLuo2000}.

\section{Numerical implementation}
\label{implementation}

\subsection{Interpolation-Supplemented Bounce-Back Boundary Conditions}
\label{interpolated_BB}

Evaluating flows past curved or non-axis-aligned geometries on a rigid Cartesian mesh necessitates boundary treatments that account for arbitrary wall placement. When a physical boundary cuts a lattice link, the boundary interface is characterized by the fractional distance $q$:
\begin{equation}
    q = \frac{|\bm{x}_{\rm f} - \bm{x}_{\rm w}|}{|\bm{x}_{\rm f} - \bm{x}_{\rm b}|}
\end{equation}
where $\bm{x}_{\rm f}$ is the fluid node, $\bm{x}_{\rm b}$ is the solid boundary node, and $\bm{x}_{\rm w}$ denotes the exact coordinates of the physical wall intersection.

\begin{figure}[htbp!]
\centering
    \includegraphics[width=0.8\textwidth]{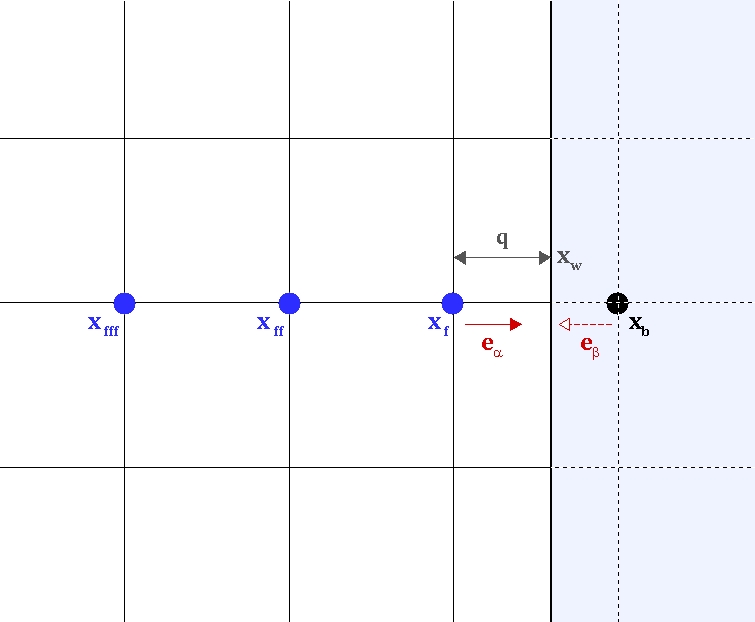}
\caption{Schematic representation of a general boundary intersecting a D2Q9 lattice link. The physical wall ($\bm{x}_{\rm w}$) is located at a fractional distance $q$ between the fluid node ($\bm{x}_{\rm f}$) and the solid boundary node ($\bm{x}_{\rm b}$). Populations stream along opposing directions defined by $\eb = -\ea$.}
\label{schematic_BB}
\end{figure}

A schematic of this structural link configuration is illustrated in Fig.~\ref{schematic_BB}. Here, $\beta$ represents the inverse velocity link direction such that $\eb=-\ea$. To preserve the algorithmic flow of the LBM, the unknown incoming distribution function traveling from the boundary node ($\bm{x}_{\rm b}$) toward the fluid node ($\bm{x}_{\rm f}$) must be evaluated prior to the streaming step. 

When $q \neq \frac{1}{2}$, standard halfway bounce-back (HWBB) conditions degrade to a geometrically crude first-order staircase approximation. To maintain second-order spatial accuracy, interpolation-supplemented bounce-back (ISBB) methods are widely employed. In the following formulations, $g^{\ast}$ denotes the post-collision PDF and $\uw$ is the local velocity vector of the solid wall.

\begin{enumerate}

\item \textbf{Bouzidi--Firdaouss--Lallemand (BFL) Model \cite{Bouzidi2001}:}

The BFL scheme achieves second-order accurate boundary tracking by utilizing a piecewise linear interpolation of the distribution functions. Depending on the value of the fractional link intersection $q$, the missing incoming population is reconstructed using the post-collision states of the nearest ($\bm{x}_{\rm f}$) and next nearest ($\bm{x}_{\rm ff}$) fluid nodes:
\begin{subequations}
\begin{align}
    g^{\ast}_{\beta}(\bm{x}_{\rm b}) &=
        2q \, g^{\ast}_\alpha(\bm{x}_{\rm f})
        + \left( 1-2q \right) g^{\ast}_\alpha(\bm{x}_{\rm ff} )
        + 2 \rho w_\alpha \left(\eb \cdot \uw \right),
        & 0 \leqslant q < \frac{1}{2}
    \\
    g^{\ast}_{\beta}(\bm{x}_{\rm b}) &=
        \frac{1}{2q} g^{\ast}_\alpha(\bm{x}_{\rm f})
        + \left( 1-\frac{1}{2q} \right) g^{\ast}_{\beta}(\bm{x}_{\rm f})
        + \frac{1}{q} \rho w_\alpha (\eb \cdot \uw),
        & \frac{1}{2} \leqslant q < 1
\label{eq_Bouzidi}
\end{align}
\end{subequations}
Because $q$ generally varies across an arbitrary surface, the asymmetric nature of these interpolation coefficients yields a non-zero net sum of local probability density across the boundary link per time step. This non-conservative numerical artifact, termed \textit{mass leakage}, stands in direct opposition to the strict mass conservation guaranteed by the symmetric halfway bounce-back method.

\item \textbf{Yu--Mei--Shyy (YMS) Model \cite{Yu2003AIAA}:}

To eliminate the piecewise programmatic discontinuity present at $q = \frac{1}{2}$ in the BFL scheme, Yu \etal \cite{Yu2003AIAA} developed a unified, single-expression boundary formulation:
\begin{equation}
    g^{\ast}_{\beta}(\bm{x}_{\rm b}) =
        \frac{q}{1+q} \left( g^{\ast}_{\beta}(\bm{x}_{\rm f})
        + g^{\ast}_\alpha(\bm{x}_{\rm f}) \right)
        + \frac{1-q}{1+q} g^{\ast}_\alpha(\bm{x}_{\rm ff})
        + \frac{2}{1+q} \rho w_\alpha (\eb \cdot \uw)
\label{eq_Yu}
\end{equation}
While this unified approach simplifies scheduling along complex surfaces, it relies on global spatial interpolations that similarly fail to conserve local mass, leaving it susceptible to identical mass leakage artifacts.

\item \textbf{Zhao--Yong (ZY) Model \cite{Zhao2017}:}

Zhao and Yong \cite{Zhao2017} introduced an alternative convex interpolation strategy requiring only a single fluid node ($\bm{x}_{\rm f}$), defined as:
\begin{equation}
    g^{\ast}_{\beta}(\bm{x}_{\rm b})
        = \frac{2q}{1+2q} g^{\ast}_{\beta}(\bm{x}_{\rm f})
        + \frac{1}{1+2q} \ga(\bm{x}_{\rm f})
        + \frac{2}{1+2q} \rho w_\alpha (\eb \cdot \uw)
\label{eq_Zhao}
\end{equation}
By omitting the secondary fluid node $\bm{x}_{\rm ff}$, this single-node formulation offers computational advantages near highly confined geometries, yet it fundamentally shares the identical mass conservation violation inherent to interpolation-supplemented conditions.

\end{enumerate}

Notably, the BFL scheme (Eq.~\eqref{eq_Bouzidi}) identically recovers the classic conservative mid-link HWBB condition \cite{Ladd1994} when $q=\frac{1}{2}$, whereas the single-expression models of Yu \etal and Zhao \& Yong do not. While all three methods offer second-order spatial accuracy and deliver equivalent hydrodynamic results, the BFL formulation is adopted here as the primary framework to demonstrate our proposed corrections.

\subsection{Momentum-Exchange Method for Force Evaluation}

A primary advantage of the LBM over conventional macroscopic solvers is its capacity to compute hydrodynamic forces directly from mesoscopic distributions without evaluating complex surface derivatives or executing explicit stress integrations. Under the conventional momentum-exchange method, the total force exerted by the fluid on a solid structure is determined by summing the net momentum transfer across all intercepted fluid-solid links \cite{Mei2002}. Enforcing Galilean invariance for moving boundaries yields the following standard relation \cite{Wen2014}:
\begin{equation}
    \mathbf{F}^{\rm \scriptscriptstyle CME} = 
        \frac{1}{c_s^2} \sum_{\bm{x}_{\rm f}} \sum_{\alpha} \left( \ea - \uw \right) g^{\ast}_{\alpha}(\bm{x}_{\rm f})
      - \frac{1}{c_s^2} \sum_{\bm{x}_{\rm b}} \sum_{\beta}  \left( \eb - \uw \right) g^{\ast}_{\beta }(\bm{x}_{\rm b})
\label{eq_ME1}
\end{equation}
where the primary summation maps over directions $\alpha$ directed from fluid nodes ($\bm{x}_{\rm f}$) toward the solid boundary, and the secondary summation loops over the corresponding conjugate directions $\beta$ directed from the solid nodes ($\bm{x}_{\rm b}$). Here, $\bm{x}_{\rm f}$ and $\bm{x}_{\rm b}$ form explicit pairs for every lattice link intersected by the physical boundary interface. Furthermore, while we utilize the notation $g^{\ast}_{\beta}(\bm{x}_{\rm b})$ for consistency with Eq.~\eqref{eq_Bouzidi}, it is worth noting that no physical collision occurs at solid boundary nodes. In practice, this term is equivalent to the post-streaming population at the fluid node $g_{\beta}(\bm{x}_{\rm f}, t+\delta_t)$.

A critical, often implied assumption in Eq.~\eqref{eq_ME1} is that the post-collision distributions $g^{\ast}_\alpha$ and $g^{\ast}_\beta$ are summed synchronously, completely independent of the true geometric position $q$. In this work, we demonstrate that this decoupled temporal treatment induces severe force artifacts when $q \neq \frac{1}{2}$. Because the physical populations travel at fixed discrete speeds, a population leaving $\bm{x}_{\rm f}$ along $\alpha$ and a population leaving $\bm{x}_{\rm b}$ along $\beta$ will cross paths precisely at the midpoint of the link ($\delta_x/2$) after exactly half a time step ($\delta_t/2$). We argue that it is only at this specific mid-link location and temporal midpoint that the opposing distribution functions must interact and exchange momentum. Evaluating the momentum exchange while ignoring the spatial mismatch when $q \neq \frac{1}{2}$ creates inaccuracy in force calculations. This issue is addressed by the modified momentum-exchange (MME) method detailed in Sec.~\ref{Poiseuille_MEM} (see Eq.~\eqref{eq_MME}).

\begin{figure}[htbp!]
\centering
    \includegraphics[width=0.5\textwidth]{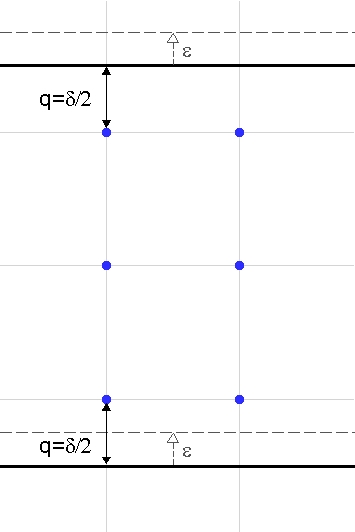}
\caption{schematic for shifting the walls.}
\label{schematic_shift}
\end{figure}

\subsection{Hydrodynamic Stress Evaluation via Distribution Functions}

To provide an independent, high-fidelity baseline for validating force evaluations, the hydrodynamic force can alternatively be determined via direct integration of the macroscopic stress tensor $\mathbf{\Pi}$. In the LBM, the non-equilibrium components of the distribution functions directly recover the viscous stress tensor algebraically, circumventing the need for noisy numerical differentiation of velocity fields:
\begin{equation}
    \mathbf{\Pi} = - \frac{\nu}{(\tau+\frac{1}{2}) c_s^4 \delta_t} \sum_\alpha \ea \ea \left( \ga-\gaeq \right)
\label{Eq_stress_tensor}
\end{equation}
The force would then be expressed as:
\begin{equation}
    \mathbf{F}^{\mbox{\tiny stress}} = \mathbf{\Pi} \cdot \mathbf{n} - p \mathbf{n}
\label{Eq_F_stress}
\end{equation}
where $\mathbf{n}$ is the unit vector normal to the boundary surface. When implemented on uniform grids, Eq.~\eqref{Eq_F_stress} evaluates the local forces directly at the regular fluid nodes ($\bm{x}_{\rm f}$ and $\bm{x}_{\rm ff}$) immediately adjacent to the boundary. To determine the true hydrodynamic force acting precisely at the physical wall ($\bm{x}_{\rm w}$), a second-order extrapolation is executed along the boundary-intercept link:
\begin{equation}
    \mathbf{F}^{\mbox{\tiny stress}}(\bm{x}_{\rm w}) = 
        (1+q) \mathbf{F}^{\mbox{\tiny stress}}(\bm{x}_{\rm f}) - q\mathbf{F}^{\mbox{\tiny stress}}(\bm{x}_{\rm ff})
\label{Eq_F_WSS}
\end{equation}

\section{Numerical Results}
\label{results}

In this section, we systematically conduct a series of benchmark simulations to illustrate the inaccuracies of the conventional momentum-exchange (CME) method for general, curved boundaries ($q \neq \frac{1}{2}$). Unless otherwise specified, the relaxation time is fixed at $\tau = 0.1$ across all simulations. We begin our validation with a simple gravity-driven Poiseuille flow and incrementally increase the geometric and kinematic complexity of the case studies.

\subsection{Poiseuille flow in a horizontal channel}
\label{Poiseuille_MEM}

Let us first consider a gravity-driven Poiseuille flow in a rectangular 2D channel. Periodic boundary conditions are applied in the $x$-direction, while no-slip conditions are imposed at the bottom and top walls. The no-slip boundaries are implemented using the widely adopted ISBB scheme outlined in Eq.~\eqref{eq_Bouzidi}. Similar results are obtained when alternative formulations, such as those in Eq.~\eqref{eq_Yu} or Eq.~\eqref{eq_Zhao}, are deployed instead.

The channel is resolved using $(N_x \times N_y) = (3 \times 10)$ lattice nodes with a Reynolds number of $\Re = \frac{\mathrm{G_x}N_y^3}{12\nu^2} = 20$, where $\mathrm{G_x}$ denotes the gravitational acceleration in the $x$-direction. Initially, the walls are positioned exactly midway between lattice links at both the bottom and top boundaries (see Fig.~\ref{schematic_shift}). Under this specific configuration ($q = \frac{1}{2}$), the drag force exerted on the top wall—calculated using the Galilean-invariant MEM \cite{Wen2014}—is expected to match the analytical solution:
\begin{equation}
    F_{a} = \frac{6 \rho \nu^2}{N_y^2} \Re
\label{eq_F_ana}
\end{equation}

\begin{figure}[htbp!]
\centering
    \includegraphics[width=0.6\textwidth]{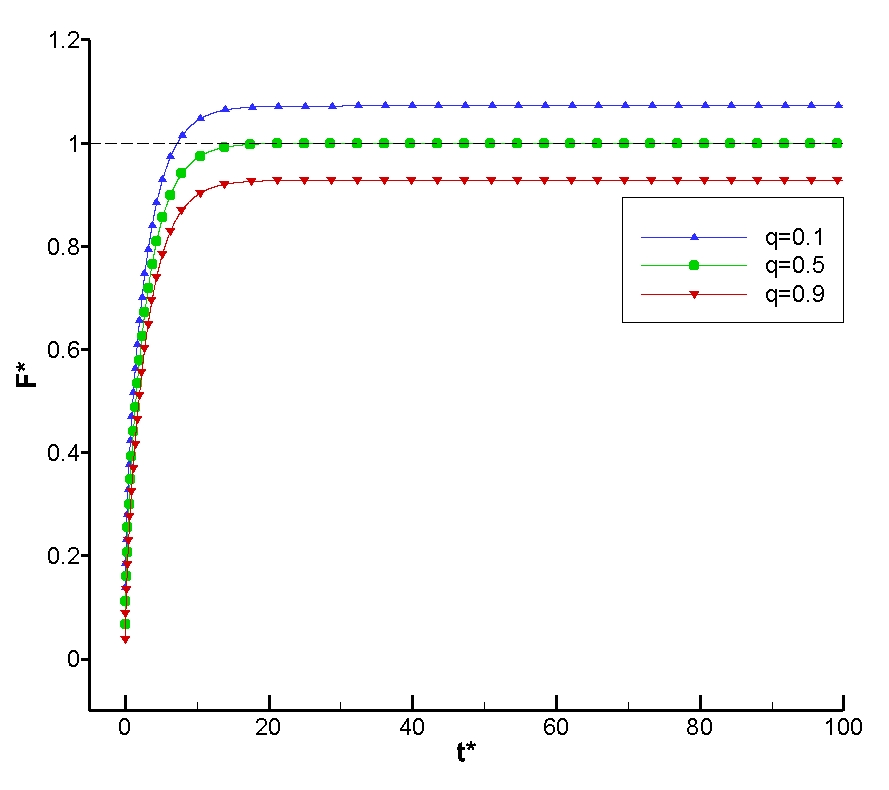}
\caption{The drag force on the top wall obtained using the original ME method (CME), illustrating the non-physical variations when $q \neq \frac{1}{2}$.}
\label{forces_org}
\end{figure}

However, when the walls are shifted by an arbitrary distance, say $\varepsilon = \pm 0.4$ lattice units, a non-negligible deviation from the analytical solution emerges, as illustrated in Fig.~\ref{forces_org}.

Let us first define the dimensionless force and time as:
\begin{subequations}
\begin{align}
    F^{\ast} &= F_{x}^ / F_{a} \\
    t^{\ast} &= \dfrac{t U_{\max}}{N_y}
\label{eq_F_t}
\end{align}
\end{subequations}
where $U_{\max}$ represents the maximum (centerline) velocity along the channel.

\begin{figure}[htbp!]
\centering
    \subfigure[]{\label{MME1} \includegraphics[width=0.8\textwidth]{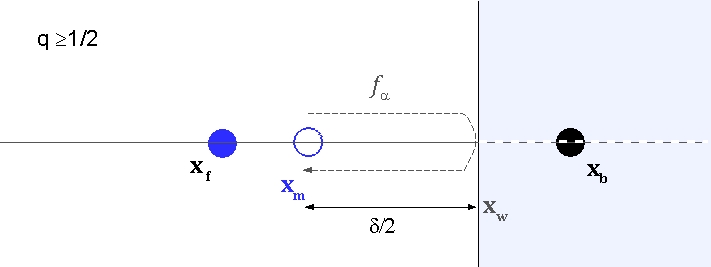}}
    \subfigure[]{\label{MME2} \includegraphics[width=0.8\textwidth]{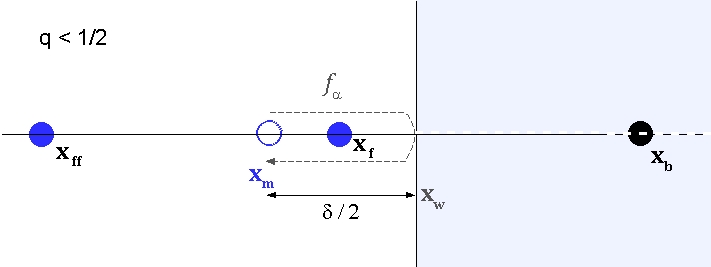}}
\caption{Schematic representation of the proposed modified momentum-exchange method under distinct geometric conditions.}
\label{schematic_MME}
\end{figure}

To rectify this issue, we argue that the interpolation procedure must explicitly ensure that the interpolated populations reach the physical wall position ($\bm{x}_{\rm w}$) exactly after half a time-step, as shown schematically in Fig.~\ref{schematic_MME}. This requirement aligns with the trapezoidal rule (or Strang splitting) used when integrating the discrete Boltzmann equation (DBE). Depending on the local value of $q$, spatial extrapolation may be required to resolve the outgoing populations at the measurement location $\bm{x}_{\rm m}$, leading to the following formulation:
\begin{equation}
    \mathbf{F}^{\rm \scriptscriptstyle MME} = 
        \frac{1}{c_s^2} \sum_{\bm{x}_{\rm m}, \alpha} \left( \ea - \uw \right) g^{\ast}_{\alpha}(\bm{x}_{\rm m},t)
      - \frac{1}{c_s^2} \sum_{\bm{x}_{\rm m}, \beta } \left( \eb - \uw \right) g_{\beta}(\bm{x}_{\rm m},t+\delta_t)
\label{eq_MME}
\end{equation}
where the post-collision distribution functions at the midpoint are evaluated via:
\begin{equation}
    g^{\ast}_{\alpha}(\bm{x}_{\rm m},t) = 
        \left(\frac{1}{2}+q\right)g^{\ast}_{\alpha}(\bm{x}_{\rm f},t)
      + \left(\frac{1}{2}-q\right)g^{\ast}_{\alpha}(\bm{x}_{\rm ff},t)
\label{eq_interpolationMME}
\end{equation}

By utilizing this refined framework, hereafter designated as the modified momentum-exchange (MME) method, the computed drag force on the upper wall perfectly aligns with the analytical solution, as demonstrated in Fig.~\ref{forces_MME_WSS}. For the halfway BB scheme, the analytical solution to the LBE~\eqref{LBE_g} is provided in Appendix~\ref{appendixA}.

\begin{figure}[htbp!]
\centering
    \subfigure[]{\label{MME} \includegraphics[width=0.48\textwidth]{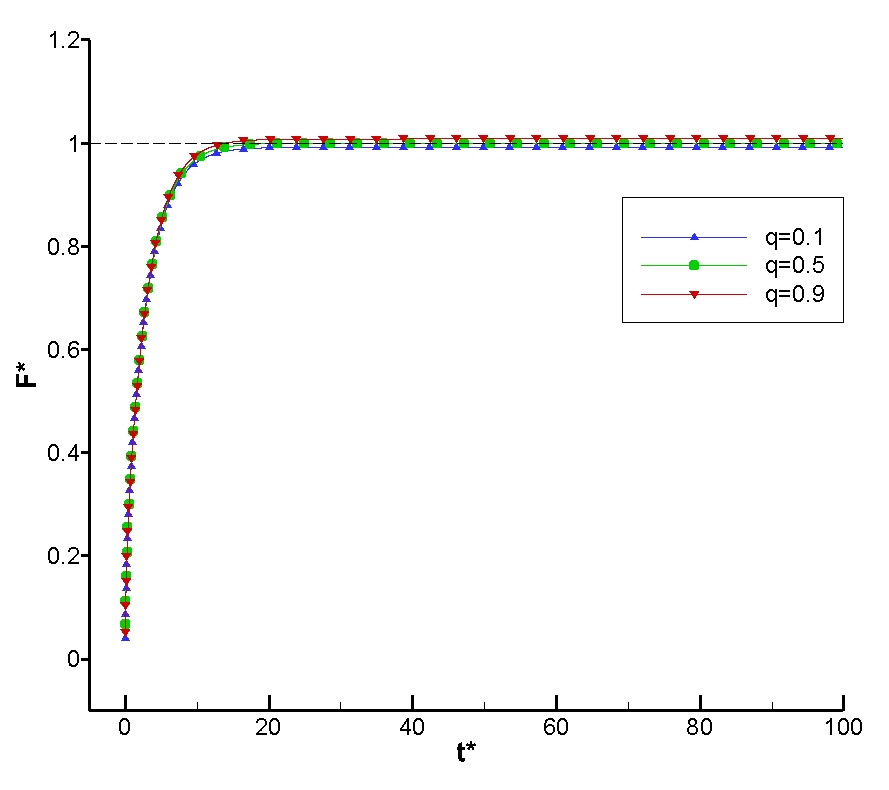}}
    \subfigure[]{\label{WSS} \includegraphics[width=0.48\textwidth]{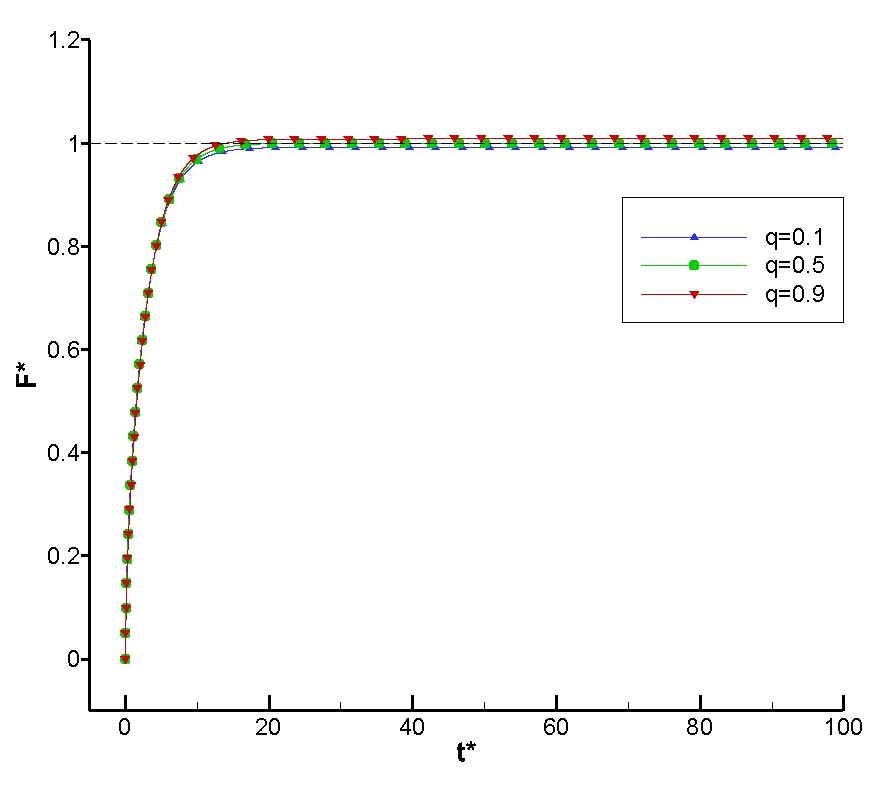}}
\caption{Drag force on the top wall evaluated using: (a) the proposed MME scheme (Eq.~\eqref{eq_MME}), and (b) direct wall shear stress integration (Eq.~\eqref{Eq_F_WSS}).}
\label{forces_MME_WSS}
\end{figure}

\begin{figure}[htbp!]
\centering
    \subfigure[]{\label{flowfield_diagonal} \includegraphics[width=0.27\textwidth]{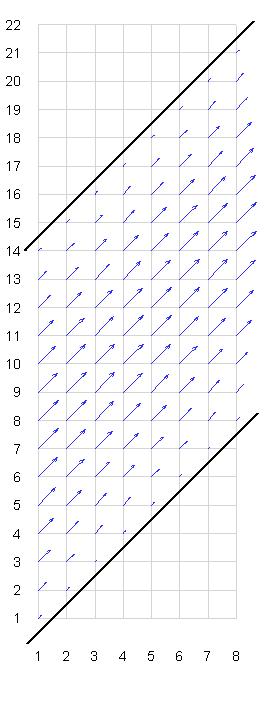}}
    \subfigure[]{\label{schematic_diagonal} \includegraphics[width=0.63\textwidth]{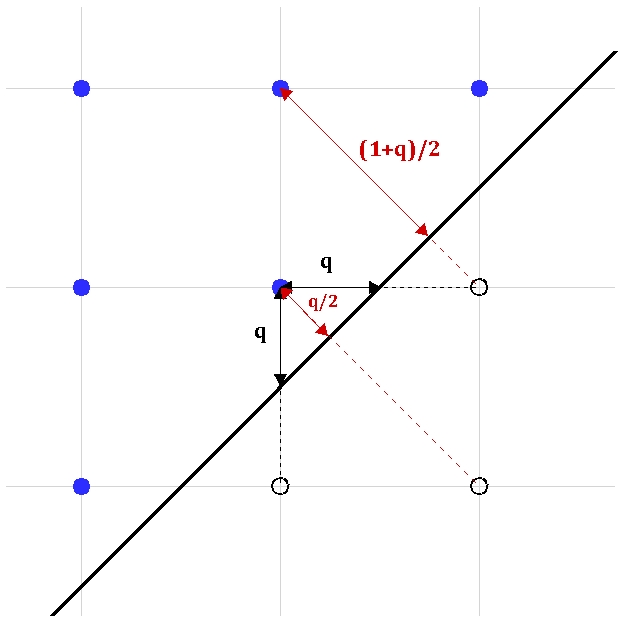}}
\caption{Poiseuille flow in a diagonally oriented channel layout: (a) simulated velocity field, and (b) schematic representation of the boundary nodes relative to the lower inclined wall.}
\label{diagonal_Poiseuille}
\end{figure}

\subsection{Poiseuille flow in an inclined channel}

Next, we examine the more subtle case of Poiseuille flow within a slanted channel. To do so, the rectangular channel is rotated by 45 degrees relative to the lattice grid, as illustrated in Fig.~\ref{diagonal_Poiseuille}, while preserving the original Reynolds number. Due to the spatial rotational invariance of the underlying physics, the drag force exerted on the upper wall should basically remain identical to the horizontal configuration.

However, in the slanted channel configuration, despite the simulated velocity profile agreeing with the analytical solution, the drag force computed via conventional methods diverges rapidly over time for all tested grid fractions ($q = 0.1, 0.5, 0.9$). A deeper probe reveals that standard ISBB formulations (Eqs.~\eqref{eq_Bouzidi}--\eqref{eq_Zhao}) inherently introduce non-physical mass-leakage at the boundaries. While this localized error does not lead to a global pressure buildup in an axis-aligned horizontal channel (cf. Fig.~\ref{forces_org}), the pressure in the diagonal system fails to reach a steady-state plateau. Similar numerical instabilities are frequently documented in conventional Navier-Stokes solvers, particularly when iterative matrix schemes are utilized without proper pressure constraints. Because the driving force in the Navier-Stokes equations is governed exclusively by the pressure gradient rather than the absolute pressure field, a classical remedy in macroscopic solvers is to pin the pressure to a reference value (typically zero) at a selected computational node.

Within the LBM framework, pinning the pressure at an arbitrary fluid cell is problematic because density and pressure are explicitly coupled via the equation of state. Alternatively, we can resolve this issue by ensuring that the boundary condition itself acts strictly mass-conservative. To achieve this, we propose modifying the rest population ($g_0$) at fluid nodes adjacent to the wall boundaries. 

For each fluid node ($\bm{x}_{\rm f}$) next to a boundary node ($\bm{x}_{\rm b}$), we track the net mass imbalance by computing the sum of the missing incoming populations from the boundary and subtracting it from the outgoing populations. This difference represents the "extra mass" artificially injected by the spatial interpolations inherent to ISBB. By subtracting this exact imbalance from the localized rest population ($\alpha = 0$), we cancel out the unphysical pressure accumulation while leaving the velocity (the first hydrodynamic moment) and the stress tensor (the second moment) completely unaltered. This correction can be seamlessly integrated into the post-collision step during the ISBB routine with minimal computational overhead:
\begin{equation}
    g^{\ast}_{0}(\bm{x}_{\rm f}, t) = g^{\ast}_{0}(\bm{x}_{\rm f}, t)
        + \sum_{\rm out} g^{\ast}_{\alpha}(\bm{x}_{\rm f}, t)
        - \sum_{\rm in}  g^{\ast}_{\beta} (\bm{x}_{\rm b}, t)
\label{eq_mass_fix}
\end{equation}
It is worth noting that for a moving wall boundary, an additional velocity-dependent term $2 \rho w_\alpha \left(\eb \cdot \uw \right)$ must be added to the right-hand side of Eq.~\eqref{eq_mass_fix} regardless of the value of $q$.

We repeat the inclined channel simulations using the mass-conservative modification proposed in Eq.~\eqref{eq_mass_fix} and evaluate the resulting drag forces. The normalized steady-state forces are summarized in Table~\ref{forces_diagonal_massConserved}. The results demonstrate that while the CME suffers from an explicit $5.3\%$ error, the proposed MME framework and the direct wall shear stress (WSS) integration yield minimal errors of only $0.4\%$ and $0.8\%$, respectively. Consequently, the force calculation error using the new MME scheme is an order of magnitude smaller than that of conventional formulations when $q \neq \frac{1}{2}$, corroborating our observations from the axis-aligned flow case.

\begin{table}[t!]
    \centering
    \begin{tabular}{c||c|c|c}
               &  CME  &  MME  &  WSS  \\
    \hline\hline
       $q=0.1$ & 1.053 & 0.996 & 0.997 \\
    \hline
       $q=0.9$ & 0.947 & 1.004 & 1.008 \\
    \end{tabular}
    \caption{Normalized drag force on the upper wall of the slanted channel utilizing the mass-conservative correction in Eq.~\eqref{eq_mass_fix}. Column results compare (a) the original momentum exchange (CME), (b) the proposed MME, and (c) the direct stress tensor integration.}
    \label{forces_diagonal_massConserved}
\end{table}

\subsection{Couette flow with a constant cross-flow velocity}

While the previous Poiseuille benchmarks featured stationary walls, we now extend our assessment to a 2D Couette flow between two parallel plates moving in opposite directions at $y=0$ ($u_w=-U_0$) and $y=N_y$ ($u_w=U_0$). In a standard setup, the steady-state velocity profile is linear: $u_x(y) = U_0 (\frac{2y}{N_y}-1)$. However, by introducing a uniform cross-flow with a constant vertical velocity $v_w=V_0$, convective acceleration distorts this profile into a non-linear exponential distribution given by:
\begin{equation}
    u_x(y) = -U_0 + 2U_0 \left(\frac{e^{\frac{y\rm{Re}}{N_y}}-1}
                                    {e^{\rm{Re}}-1} \right)
\label{eq_Couette_u}
\end{equation}
where $\rm{Re} = \frac{V_0 N_y}{\nu}$ represents the cross-flow Reynolds number. The corresponding horizontal force per unit length exerted by the fluid on the top wall is analytically derived as:
\begin{equation}
    F_{w} = -2\rho U_0 V_0 \left( \frac{e^{\rm{Re}}}{e^{\rm{Re}}-1} \right)
\label{eq_Couette_F}
\end{equation}
For this particular benchmark, both momentum exchange methods accurately recover the analytical solution across all the considered variations of the grid boundary fraction $q$.
The identical performance of the CME and the proposed MME in this benchmark could be due to the unique decoupling of the horizontal momentum transfer from the sub-grid wall position. In this configuration, the uniform cross-flow velocity $v_w = V_0$ establishes a constant convective mass flux across the channel height. At steady state, because the horizontal velocity gradient $\partial u_x / \partial y$ at the wall is governed entirely by the cross-flow, the local spatial variations induced by shifting the wall boundary within a lattice link ($q \neq \frac{1}{2}$) do not alter the net horizontal momentum exchange between the fluid and the plates. Consequently, the sub-grid geometric corrections of the MME scheme yield no additional optimization over the standard MEM when the force calculation is dominated by a uniform, continuous convective flux normal to the boundary.

\subsection{Taylor-Couette flow}

\begin{figure}[htbp!]
\centering
    \includegraphics[width=0.6\textwidth]{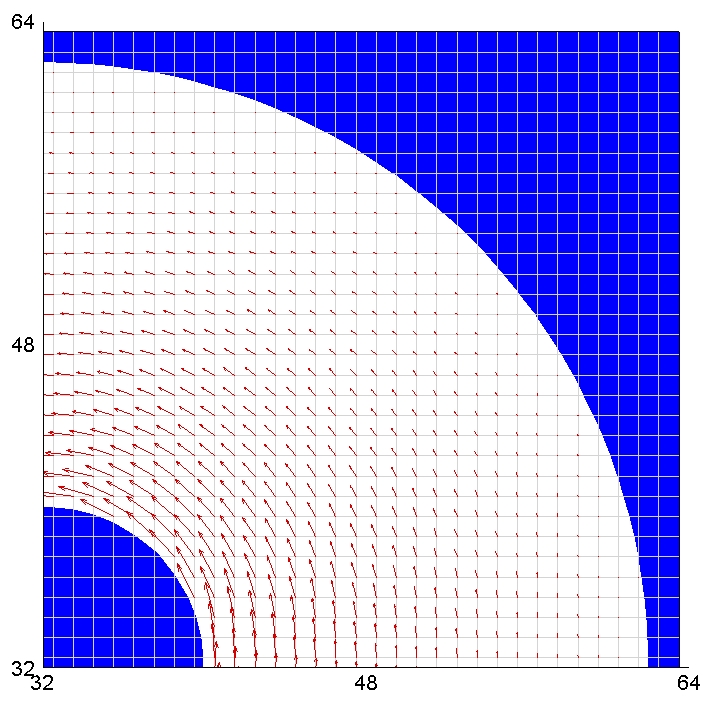}
\caption{A quarter of the geometrical setup in the cylindrical Couette flow.}
\label{Cylindrical_Couette}
\end{figure}

To further challenge the proposed enhancements against curved geometries, we evaluate a Taylor-Couette flow confined between two concentric cylinders with radii $R_1$ and $R_2$. The inner cylinder rotates at a constant angular velocity $\omega_1$, while the outer cylinder remains stationary. The computational domain is defined on an $N_x \times N_x$ lattice, where the radii are set to $R_1=N_x/8$ and $R_2=3N_x/8$. To evaluate the grid convergence and dependency behavior of our scheme, the grid resolution is scaled using $N_x = 16 h$, where the scaling parameter is varied as $h=1,2,4,8$. A quarter-domain representation of this geometrical setup is depicted in Fig.~\ref{Cylindrical_Couette}.

The analytical profiles for the steady azimuthal velocity and radial pressure are given by:
\begin{subequations}
\begin{align}
    U_{\theta}^{\rm ana}(r) &= - r \omega_1 \frac{R_1^2}{R_2^2-R_1^2}
                            \left(1-\frac{R_2^2}{r^2} \right)
    \\
    p^{\rm ana}(r) &= \frac{1}{2}\rho \omega_1^2 \frac{R_1^4}{\left(R_2^2-R_1^2\right)^2}
                \left( 4R_2^2 \ln\left(\frac{R_2}{r}\right) + \frac{r^4-R_2^4}{r^2} \right)
\label{eq_u_p}
\end{align}
\end{subequations}
The net torque can be evaluated via $\mathbf{T}=\mathbf{r} \times \mathbf{F}$, where the force vector $\mathbf{F}$ is computed using either the conventional ME (CME) or the proposed MME method. The analytical torque per unit length acting on the inner cylinder is given by:
\begin{equation}
    \mathbf{T} = 4\pi \mu \omega_1 \frac{R_1^2R_2^2}{R_2^2-R_1^2} \bm{\hat{k}}
\label{eq_torq}
\end{equation}

\begin{figure}[htbp!]
\centering
    \subfigure[]{\includegraphics[width=0.49\textwidth]{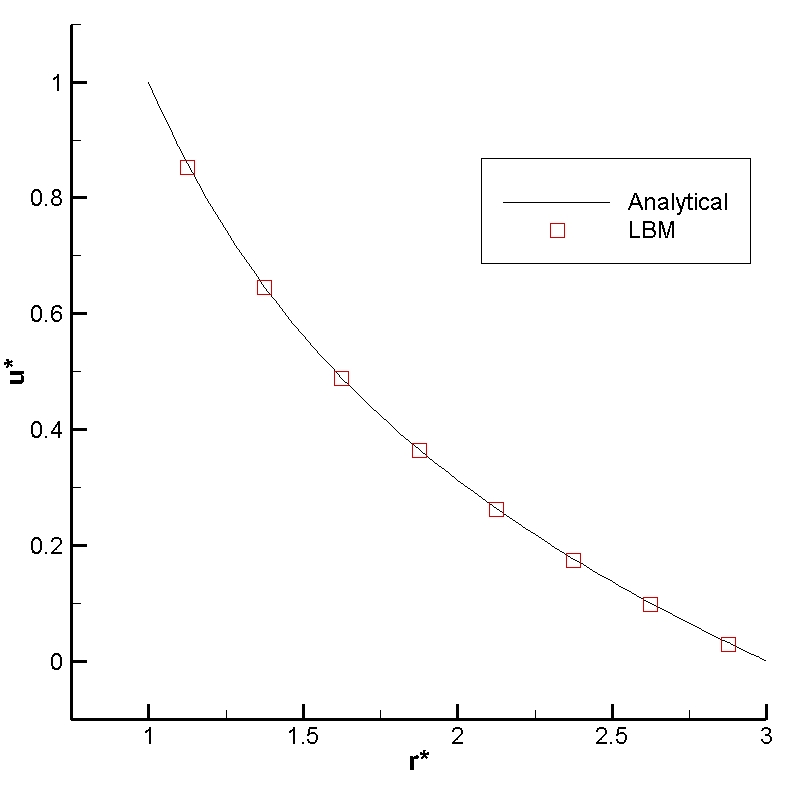}}
    \subfigure[]{\includegraphics[width=0.49\textwidth]{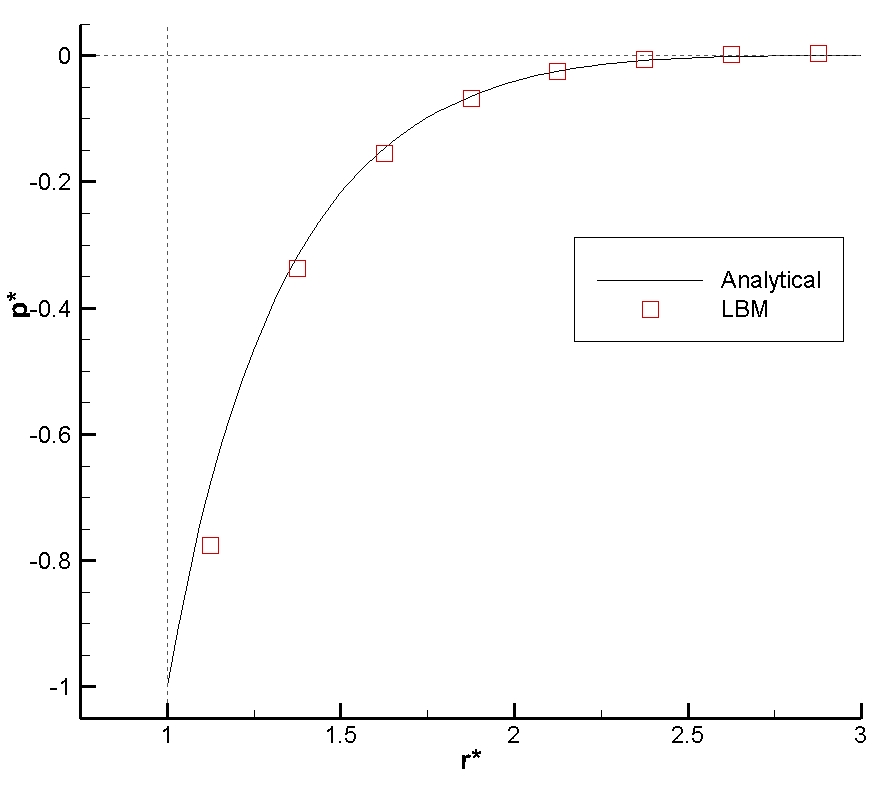}}
\caption{Steady state solution for the Cylindrical Couette flow. (a) velocity profile; (b) pressure profile.}
\label{rotary_profiles}
\end{figure}

To maximize numerical accuracy and minimize initial startup transients, the pressure field is initialized to match the analytical distribution. The steady-state velocity and pressure profiles, normalized by their analytical counterparts evaluated at $r=R_1$, are shown in Fig.~\ref{rotary_profiles}.

The computed torque on the inner cylinder, normalized against the analytical value from Eq.~\eqref{eq_torq}, is provided in Table~\ref{torque}.
Interestingly, the data in Table~\ref{torque} reveals that the CME yields highly accurate torque predictions even at low grid resolutions ($h=1,2$), matching or slightly outperforming the MME scheme before both converge at higher resolutions ($h=8$). This behavior might be attributed to a statistical error-cancellation effect inherent to continuously curved, axis-symmetric geometries mapped onto a uniform Cartesian grid. For a circular cylinder, the sub-grid boundary fractions $q$ are essentially randomly distributed across the circumferential nodes, spanning the entire range from $0$ to $1$. When computing the global torque, the overestimations of force at nodes where $q > \frac{1}{2}$ are counterbalanced by underestimations at nodes where $q < \frac{1}{2}$. This localized integration smoothing mimics second-order accuracy globally, meaning the node-by-node geometric corrections introduced by the MME scheme provide minimal visible improvement when integrated over a closed, symmetric curved surface.

\begin{table}[t!]
    \centering
    \begin{tabular}{c||c|c|c|c}
        $h$ & 1    & 2    & 4    & 8    \\
    \hline\hline
        CME & 0.95 & 1.00 & 1.01 & 1.00 \\
    \hline
        MME & 0.90 & 1.03 & 1.07 & 1.02
    \end{tabular}
    \caption{Normalized torque exerted on the rotating cylinder, comparing the conventional ME and the proposed MME method across different grid resolution scaling factors $h$.}
    \label{torque}
\end{table}

\section{Summary and conclusion}
\label{conclusion}

In this work, we have systematically investigated and resolved two distinct, yet interconnected, sources of numerical error that plague hydrodynamic force and pressure calculations in lattice Boltzmann simulations involving curved boundaries: the geometric limitation of the conventional momentum-exchange (CME) method when $q \neq \frac{1}{2}$, and the non-physical mass leakage inherent to interpolation-supplemented bounce-back (ISBB) treatments. 

To address the geometric discrepancy, we introduced the Modified Momentum-Exchange (MME) method. By explicitly accounting for the sub-grid boundary fraction $q$ via spatial interpolation/extrapolation, the MME framework ensures that the fluid populations used to calculate hydrodynamic forces reach the physical wall boundary exactly at the half-time-step interval. This alignment structurally satisfies the trapezoidal rule underlying the integration of the discrete Boltzmann equation. The utility of the MME scheme was validated using axis-aligned Poiseuille flow and Couette cross-flow benchmarks, where it successfully eliminated the non-physical force oscillations observed in the CME when the solid wall is shifted off the mid-link position.

Furthermore, we probed a critical vulnerability in standard ISBB formulations: an inherent violation of local mass conservation due to spatial interpolation. While benign in simple axis-aligned channels, this mass leakage induces severe, unbounded pressure accumulation in complex or rotated geometries—as demonstrated by our 45-degree inclined Poiseuille flow benchmark. To rectify this instability, we proposed a computationally elegant, mass-conservative remedy. By tracking the localized net mass imbalance generated during the ISBB step and subtracting it directly from the rest distribution function of the adjacent fluid nodes, we effectively neutralized the unphysical pressure buildup. Crucially, because this correction isolates the mass imbalance within the zeroth hydrodynamic moment, it eliminates pressure divergence without altering the local fluid velocity or stress tensor fields.

When combined, the mass-conservative ISBB framework and the MME scheme demonstrate improved accuracy and robust stability. In the inclined channel benchmark, the proposed framework reduced force calculation errors by an order of magnitude compared to the CME, yielding excellent agreement with direct wall shear stress integration. Finally, application to the Taylor-Couette flow demonstrated the scheme's adaptability to curved geometries. 

In conclusion, the proposed modifications offer a highly accurate, mass-conservative, and computationally efficient framework for modeling fluid-structure interactions on curved boundaries. Because the corrections are implemented entirely during the post-collision and streaming phases with minimal computational overhead, they can be seamlessly integrated into existing LBM codes. Although extension of the proposed MME framework to three-dimensional complex geometries is trivial, future work will focus on moving boundary problems involving fluid-particle suspensions.

\section*{Acknowledgment}

The author is grateful to Dr. Cheng Peng for helpful discussions.

\appendix
\section{Analytical solution}
\label{appendixA}

The solution to the Navier-Stokes equation for a gravity-driven Poiseuille flow gives us the following parabolic velocity profile across the channel height, $L_y$:
\begin{equation}
    u_x(j) = \dfrac{\mathrm{G_x}}{2 \nu} \left( j-0.5 \right) \left( N-j+0.5 \right); \quad j = 1, 2, ..., N
\end{equation}

For simplicity, We consider the LBE~\eqref{LBE_g} with a BGK collision. Taking advantage of the periodicity in the x-direction, we can obtain the following equations:
\begin{alignat*}{4}
    &g_0(j) &=& {}
            &+& \dfrac{4}{9}\left[ p-\dfrac{1}{2}\rho u_x^2(j) \right]
            &-& \dfrac{4\delta_t}{9} \dfrac{2\tau-1}{2} \rho \mathrm{G_x} u_x(j) \\
    &g_1(j) &=& {}
            &+& \dfrac{1}{9}\left[ p +\rho u_x(j) + \rho u_x^2(j) \right]
            &+& \dfrac{ \delta_t}{9} \dfrac{2\tau-1}{2} \rho \mathrm{G_x} [1+2u_x(j)] \\
    &g_2(j+1) &=& \dfrac{\tau-1}{\tau}g_2(j)
            &+& \dfrac{1}{9\tau}\left[ p-\dfrac{1}{2}\rho u_x^2(j) \right]
            &-& \dfrac{ \delta_t}{9} \dfrac{2\tau-1}{2\tau} \rho \mathrm{G_x} u_x(j) \\
    &g_3(j) &=& {}
            &+& \dfrac{1}{9}\left[ p -\rho u_x(j) + \rho u_x^2(j) \right]
            &-& \dfrac{ \delta_t}{9} \dfrac{2\tau-1}{2} \rho \mathrm{G_x} [1-2u(j)] \\
    &g_4(j-1) &=& \dfrac{\tau-1}{\tau}g_4(j)
            &+& \dfrac{1}{9\tau}\left[ p-\dfrac{1}{2}\rho u_x^2(j) \right]
            &-& \dfrac{ \delta_t}{9} \dfrac{2\tau-1}{2\tau} \rho \mathrm{G_x} u_x(j) \\
    &g_5(j+1) &=& \dfrac{\tau-1}{\tau}g_5(j)
            &+& \dfrac{1}{36\tau}\left[ p + \rho u_x(j) + \rho u_x^2(j) \right]
            &+& \dfrac{\delta_t}{36} \dfrac{2\tau-1}{2\tau} \rho \mathrm{G_x} [1+2u(j)] \\
    &g_6(j+1) &=& \dfrac{\tau-1}{\tau}g_6(j)
            &+& \dfrac{1}{36\tau}\left[ p - \rho u_x(j) + \rho u_x^2(j) \right]
            &-& \dfrac{\delta_t}{36} \dfrac{2\tau-1}{2\tau} \rho \mathrm{G_x} [1-2u(j)] \\
    &g_7(j-1) &=& \dfrac{\tau-1}{\tau}g_7(j) 
            &+& \dfrac{1}{36\tau}\left[ p - \rho u_x(j) + \rho u_x^2(j) \right] 
            &-& \dfrac{\delta_t}{36} \dfrac{2\tau-1}{2\tau} \rho \mathrm{G_x} [1-2u(j)] \\
    &g_8(j-1) &=& \dfrac{\tau-1}{\tau}g_8(j) 
            &+& \dfrac{1}{36\tau}\left[ p + \rho u_x(j) + \rho u_x^2(j) \right] 
            &+& \dfrac{\delta_t}{36} \dfrac{2\tau-1}{2\tau} \rho \mathrm{G_x} [1+2u(j)]
\end{alignat*}
On the other hand, the half-way BB boundary conditions translate to the following relations at $j = 1$ and $j = N$:
\begin{equation}
    \begin{cases}
        g_2(1) = g_4(0) \\
        g_5(1) = g_7(0) \\
        g_6(1) = g_8(0)
    \end{cases}
\end{equation}
and
\begin{equation}
    \begin{cases}
        g_4(N) = g_2(N+1) \\
        g_7(N) = g_5(N+1) \\
        g_8(N) = g_6(N+1)
    \end{cases}
\end{equation}
We can solve the above system of equations for each pair along the bounce-back links; for example, for $g_5$ and $g_7$:
\setcounter{MaxMatrixCols}{11} 
\[
\setlength{\arraycolsep}{3pt} 
\begin{pmatrix}
    \tau   &1-\tau &       &        &        &        &        &        &       &       &        \\
    1-\tau &  0    & \tau  &        &        &        &        &        &       &       &        \\
           & \tau  &  0    & 1-\tau &        &        &        &        &       &       &        \\
           &       &1-\tau &  0     & \tau   &        &        &        &       &       &        \\
           &       &       & \tau   &  0     & 1-\tau &        &        &       &       &        \\
           &       &       & \ddots & \ddots & \ddots & \ddots & \ddots &       &       &        \\
           &       &       &        &        & 1-\tau &  0     & \tau   &       &       &        \\
           &       &       &        &        &        & \tau   &  0     &1-\tau &       &        \\
           &       &       &        &        &        &        &1-\tau  &  0    & \tau  &        \\
           &       &       &        &        &        &        &        & \tau  &  0    & 1-\tau \\
           &       &       &        &        &        &        &        &       &1-\tau & \tau
\end{pmatrix}
\begin{pmatrix}
    g_5(1) \\ g_7(1) \\ g_5(2) \\ g_7(2) \\ g_5(3) \\
    \vdots \\
    g_7(N-2) \\ g_5(N-1) \\ g_7(N-1) \\ g_5(N) \\ g_7(N)
\end{pmatrix}
=
\begin{bmatrix}
    R_{68}(1) \\ R_{57}(1) \\ R_{68}(2) \\ R_{57}(2) \\ R_{68}(3) \\
    \vdots \\
    R_{57}(N-2) \\ R_{68}(N-1) \\ R_{57}(N-1) \\ R_{68}(N) \\ R_{57}(N)
\end{bmatrix}
\]
%
%
where
\begin{eqnarray*}
    R_{57}(j) = \dfrac{1}{36}\left[ p - \rho u_x(j) + \rho u_x^2(j) - 3\rho\nu \mathrm{G_x} (1-2u(j)) \right] \\
    R_{68}(j) = \dfrac{1}{36}\left[ p + \rho u_x(j) + \rho u_x^2(j) + 3\rho\nu \mathrm{G_x} (1+2u(j)) \right]
\end{eqnarray*}

Since the general solution to the above system is non-trivial, we showcase the solution for N=3 and N=5. Of interest is the second-order moments that appear in the calculation of shear stress in Eq.~\eqref{Eq_stress_tensor}. For the off-diagonal component of the shear stress, the analytical solution gives us the following:
for N=3:
\begin{equation}
\sum_\alpha ( \ga-\gaeq ) e_{\alpha x} e_{\alpha y} = \left\{
\begin{aligned}
        & \frac{(1+\lambda^3)[\Delta(1)+\Delta(2)]}
        {1+\lambda+\lambda^2+\lambda^3+\lambda^4+\lambda^5}, & j=1 \\
        & 0, & j=2 \\
        -&\frac{(1+\lambda^3)[\Delta(2)+\Delta(3)]}
        {1+\lambda+\lambda^2+\lambda^3+\lambda^4+\lambda^5}, & j=3
\end{aligned}
\right.
\end{equation}
and for N=5:
\begin{equation}
\sum_\alpha ( \ga-\gaeq ) e_{\alpha x} e_{\alpha y} = 
\begin{cases}
    \frac{1}{1-\lambda^{10}}
    \left[
    \begin{aligned}
        -(1-\lambda) r^3 (1+\lambda  ) \Delta(1) \\
        +(1-\lambda) r^2 (1+\lambda^3) \Delta(2) \\
        -(1-\lambda) r   (1+\lambda^5) \Delta(3) \\
        +(1-\lambda)     (1+\lambda^7) \Delta(4) \\
        +                (1-\lambda^8) \Delta(5)
    \end{aligned}
    \right], & j=1 \\[2.5ex]
    
    \frac{1}{1-\lambda^{10}}
    \left[
    \begin{aligned}
         (1-\lambda^3) r^2 (1+\lambda  ) \Delta(1) \\
        +(1-\lambda^3) r   (1+\lambda^3) \Delta(2) \\
        -(1-\lambda^3)     (1+\lambda^5) \Delta(3) \\
        +              r^2 (1-\lambda^4) \Delta(4) \\
        +(1-\lambda^7)     (1+\lambda  ) \Delta(5)
    \end{aligned}
    \right], & j=2 \\[2.5ex]
    
    0, & j=3 \\[1.5ex]
    
    \frac{1}{1-\lambda^{10}}
    \left[
    \begin{aligned}
        -(1-\lambda^3) r^2 (1+\lambda  ) \Delta(1) \\
        -(1-\lambda^3) r   (1+\lambda^3) \Delta(2) \\
        +(1-\lambda^3)     (1+\lambda^5) \Delta(3) \\
        -              r^2 (1-\lambda^4) \Delta(4) \\
        -(1-\lambda^7)     (1+\lambda  ) \Delta(5)
    \end{aligned}
    \right], & j=4 \\[2.5ex]
    
    \frac{1}{1-\lambda^{10}}
    \left[
    \begin{aligned}
         (1-\lambda) r^3 (1+\lambda  ) \Delta(1) \\
        -(1-\lambda) r^2 (1+\lambda^3) \Delta(2) \\
        +(1-\lambda) r   (1+\lambda^5) \Delta(3) \\
        -(1-\lambda)     (1+\lambda^7) \Delta(4) \\
        -                (1-\lambda^8) \Delta(5)
    \end{aligned}
    \right], & j=5
\end{cases}
\end{equation}
where
\begin{equation}
    \lambda=\dfrac{1-\tau}{\tau}
\end{equation}
and
\begin{equation}
    \Delta(j) = \frac{\rho}{18\tau} \left[ u_x(j) + 3\nu \mathrm{G_x} \right]
\end{equation}


\newpage

\bibliography{Refrences.bib}

\end{document}